\documentclass[11pt]{article}

\usepackage[utf8]{inputenc}
\usepackage[vmargin=1.25in]{geometry}
\usepackage{amsmath,amssymb}
\usepackage{mathrsfs}
\usepackage[square]{natbib}
\usepackage{graphicx}
\usepackage[colorlinks,citecolor=green,linkcolor=blue,bookmarks=false,hypertexnames=true]{hyperref}
\usepackage{stackengine}

\newtheorem{remark}{Remark}

\def\sar{\emph{Sargassum}}

\newcommand{\defn}[1]{\emph{#1}}
\newcommand\restr[2]{
  \left.\kern-\nulldelimiterspace 
  #1 
  \right|_{#2} 
}
\renewcommand{\Pr}{\ensuremath{\operatorname{prob}}}

\usepackage{titling}
\usepackage{array}
\preauthor{\begin{center}
    \large \lineskip .75em%
        \begin{tabular}[t]{>{\centering\arraybackslash}p{.45\textwidth}}}
        \postauthor{\end{tabular}\par\end{center}}
\makeatletter
\renewcommand\and{ 
  \end{tabular}%
  \hfill
  \begin{tabular}[t]{>{\centering\arraybackslash}p{.45\textwidth}}} 
\makeatother

\begin{document}

\title{Dynamical geography and transition paths of \sar{ }in
the tropical Atlantic}

\author{F.J.\ Beron-Vera\thanks{Corresponding author.}\\ Department
of Atmospheric Sciences\\ Rosenstiel School of Marine \& Atmospheric
Science\\ University of Miami\\ Miami, FL 33149 USA\\ {\small
fberon@miami.edu} \and M.J.\ Olascoaga\\ Department of Ocean
Sciences\\ Rosenstiel School of Marine \& Atmospheric Science\\
University of Miami\\ Miami, FL 33149 USA\\ {\small
jolascoaga@miami.edu}\and N.F.\ Putman\\LGL Ecological Research
Assoc.\ Inc.\\ Bryan, TX 77802 USA\\ {\small nathan.putman@gmail.com}\and
J.\ Tri\~nanes\thanks{Also at Cooperative Institute for Marine \&
Atmospheric Studies, University of Miami, Miami, Florida, USA and
Departamento de Electr\'onica y Computaci\'on, Universidade de
Santiago de Compostela, Santiago, Spain.}\\ Atlantic Oceanographic
and Meteorological Laboratory\\ National Oceanic \& Atmospheric
Administration\\ Miami, FL 33149 USA\\ {\small
joaquin.trinanes@noaa.gov}\and G.J.\ Goni\\Atlantic Oceanographic
and Meteorological Laboratory\\ National Oceanic \& Atmospheric
Administration\\ Miami, FL 33149 USA\\ {\small gustavo.goni@noaa.gov}\and
R.\ Lumpkin\\Atlantic Oceanographic and Meteorological Laboratory\\
National Oceanic \& Atmospheric Administration\\ Miami, FL 33149
USA\\ {\small rick.lumpkin@noaa.gov}
}

\date{Started: February 28, 2021. This version: \today. To
appear in \emph{AIP Advences}.\vspace{-0.25in}}

\maketitle

\begin{abstract}
  By analyzing a time-homogeneous Markov chain constructed using
  trajectories of undrogued drifting buoys from the NOAA's Global
  Drifter Program, we find that probability density can distribute
  in a manner that resembles very closely the recently observed
  recurrent belt of high \sar{ }density in the tropical Atlantic
  between 5--10$^{\circ}$N, coined the \emph{Great Atlantic \sar{
  }Belt} (\emph{GASB}). A spectral analysis of the associated
  transition matrix further unveils a forward attracting almost-invariant
  set in the northwestern Gulf of Mexico with a corresponding basin
  of attraction disconnected from the Sargasso Sea, but including
  the nutrient-rich regions around the Amazon and Orinoco rivers
  mouths and also the upwelling system off the northern coast of
  west Africa. This represents a data-based inference of potential
  remote sources of \sar{ }recurrently invading the Intra-Americas
  Seas (IAS). By further applying Transition Path Theory (TPT) on
  the data-derived Markov chain model, two potential pathways for
  \sar{ }into the IAS from the upwelling system off the coast of
  Africa are revealed.  One TPT-inferred pathway takes place along
  the GASB.  The second pathway is more southern and slower, first
  going through the Gulf of Guinea, then across the tropical Atlantic
  toward the mouth of the Amazon River, and finally along the
  northeastern South American margin.  The existence of such a
  southern TPT-inferred pathway may have consequences for bloom
  stimulation by nutrients from river runoff.

\end{abstract}

\section{Introduction}

Since the early 2010s, pelagic \emph{Sargassum}, a genus of brown
macroalgae that forms floating rafts at the ocean surface, has
inundated the Intra-American Seas (IAS), particularly the Caribbean
Sea, during spring and summer months \citep{Wang-etal-19}.  These
rafts of algae serve as important habitats for diverse marine fauna
\citep{Bertola-etal-20} and have been argued to be important carbon
sinks \citep{Paraguay-etal-20}.  Nonetheless, these rafts carry
large amounts of toxic substances and heavy metals and when they
enter coastal zones they can result in mortality to fishes and sea
turtles and effectively smother seagrass and coral communities
\citep{vanTussenbroek-etal-17}. As large \emph{Sargassum} rafts
decompose onshore they give off toxic amounts of hydrogen sulphide
that can cause health problems in humans, the volume that washes
ashore on beaches also diminishes tourism and, as a result, disrupts
the local economy \citep{Smetacek-Zingone-13, Resiere-etal-18}.

\begin{figure}[p!]
  \centering%
  \includegraphics[width=.65\textwidth]{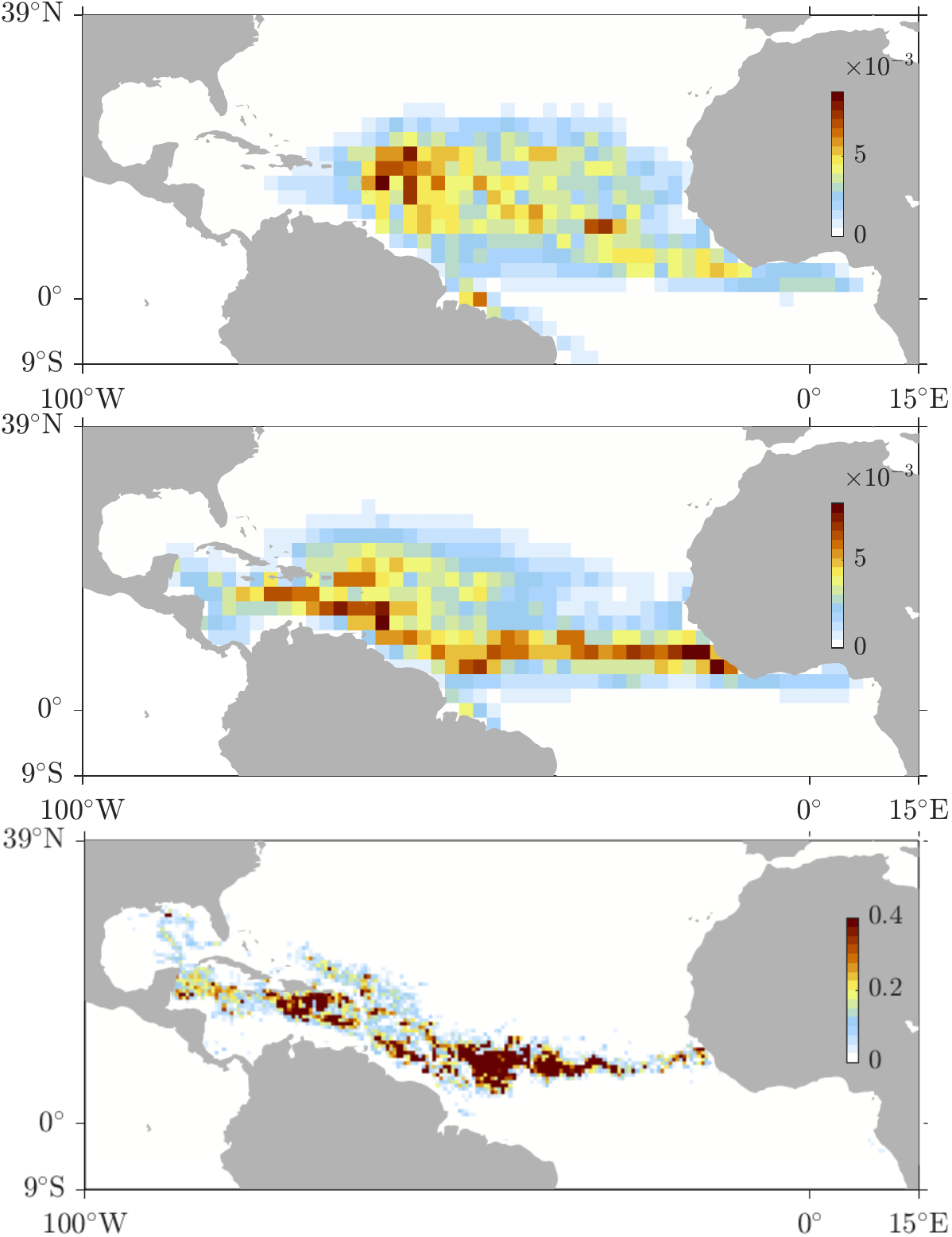}%
  \caption{Starting from a uniform distribution, discrete probability
  density of finding tracer after four months of evolution along a
  Markov chain constructed based on drogued (top panel) and undrogued
  (middle panel) drifter trajectories from the NOAA Global Drifter
  Program (GDP).  (bottom panel) The \emph{Great Atlantic Sargassum
  belt} (\emph{GASB}) as inferred in 2015 from the NASA Moderate
  Resolution Imaging Spectroradiometer (MODIS) aboard \emph{Terra}
  and \emph{Aqua} satellites.  The quantity plotted is percentage
  of \sar{ }coverage per pixel as determined by the Alternative
  Floating Algae Index (AFAI), a measure of the magnitude of red
  edge reflectance of floating \sar.}
  \label{fig:gasb}%
\end{figure}

Comprehensive analyses of satellite imagery across the Atlantic
Ocean, revealed a recurrent band of \sar{} high density between
about 5 and 10$^\circ$N, referred to as the \emph{Great Atlantic
Sargassum Belt} (\emph{GASB}) in \citep{Wang-etal-19},
often extending off the coast of West Africa to the Gulf of Mexico
(Fig.\ \ref{fig:gasb}, bottom panel). The factors that precipitated
blooms of pelagic \sar{ }in the tropical Atlantic remain an area
of active debate, as do the factors that maintain its occurrence
across this region \citep{Jouanno-etal-21a, Jouanno-etal-21b,
LaPointe-etal-21}.  Several published studies have attempted to
account for the extreme spatiotemporal variability in its distribution
by simulating the movement of \sar{ }rafts as passive tracers
advected by surface velocity fields produced by ocean circulation
models.  Conclusions about the connectivity between \sar{ }blooms
in the tropical Atlantic and the Sargasso Sea fundamentally differ
depending whether simulations assume winds contribute to \sar{
}movement \citep{Franks-etal-16, Wang-etal-19, Johns-etal-20}.
Available empirical data suggest that accounting for windage effects
improve predictions of \sar{ }raft trajectories \citep{Putman-etal-20}
and distribution \citep{Berline-etal-20}.

The importance of windage in modeling \sar{ }movement is demonstrated
in the top and middle panels of Fig.\ \ref{fig:gasb}, which show
discrete probability densities of finding \sar{ }in the tropical
Atlantic after three months of evolution from an initially uniform
distribution inside $\mathscr I = $ [5$^{\circ}$S, 18$^{\circ}$N]
$\times$ [58$^{\circ}$W, 15$^{\circ}$W] as in \citep{Wang-etal-19}.
Unlike \citep{Wang-etal-19} and earlier work, which resorted to
tracer advection by simulated velocity, the evolution is here
provided by a transfer operator \citep{Lasota-Mackey-94} obtained
by discretizing \citep{Ulam-60} the motion as described by the
trajectories $x(t)$ of surface drifting buoys from the National
Oceanic and Atmospheric Administration (NOAA) Global Drifter Program
(GDP) \citep{Lumpkin-Pazos-07} assuming that the underlying dynamics
represent a passive advection--diffusion process (e.g.,
\citep{Miron-etal-19-JPO}).

The discrete transfer operator is given by a \emph{transition matrix}
$\mathsf P = (P_{ij}) \in \mathbb R^{N\times N}$ of conditional
probabilities of $x(t)$ to moving between nonintersecting boxes
$\{b_1,\dotsc, b_N\}$ covering the surface of the Atlantic between
9$^{\circ}$S and 39$^{\circ}$N, denoted $\mathscr D$, viz.,
\begin{subequations}
\begin{equation}
  P_{ij} = \Pr(X_{n+1}\in b_j\mid X_n\in b_i) \approx
  \frac{C_{ij}}{\sum_{k=1}^N C_{ik}},
\end{equation}
where
\begin{equation}
  C_{ij} := \#\big\{x(t)\in b_i,\, x(t+T) \in
  b_j\big\}
\end{equation}
\label{eq:P}%
\end{subequations}
independent of the start time $t$.  Here $X_n$ denotes random
position at discrete time $nT$, where $T$ is the \emph{transition
time step}.  The boxes represent the states of a \defn{Markov chain},
namely, a stochastic model describing the stochastic state transitions
in which the transition probability of each state depends only on
the state attained in the previous event \citep{Norris-98}.  By
ignoring the start time $t$ of a trajectory in \eqref{eq:P}, the
chain is rendered time-homogeneous.  The evolution of $\Pr(X_n\in
b_i)$ is obtained under left multiplication by $\mathsf P$, i.e.,
\begin{equation}
  \Pr(X_{n+1}\in b_j) = \sum_{i=1}^N P_{ij} \Pr(X_n\in b_i).
\end{equation}
Figure \ref{fig:gasb} specifically shows $\Pr(X_{24} \in b_i)$,
where $\Pr(X_0\in b_i) = 1/|I|$ for $b_{i\in I}$ spanning $\mathscr
I\subset \mathscr D$ and 0 otherwise. Assuming $T = 5$\,d, as we do
here, this represents the end of nearly four months of evolution.
Our transition time choice $T = 5$\,d is long enough for Markovian
dynamics to approximately hold, as it is longer than a typical
Lagrangian integral time scale of about 1 day at the ocean surface
\citep{LaCasce-08} and as noted in pioneering \citep{Maximenko-etal-12}
and most recent \citep{Miron-etal-17, Olascoaga-etal-18,
McAdam-vanSebille-18, Miron-etal-19-Chaos, Miron-etal-19-JPO,
Miron-etal-21-Chaos, Drouin-etal-22} work involving observed, rather
than simulated, trajectory data.

A critical difference between the top and middle panels of Fig.\
\ref{fig:gasb} is that the former uses trajectories of drifters
drogued at 15 m, while the latter uses trajectories of undrogued
drifters \citep{Lumpkin-etal-12}.  Note that the assessment based
on undrogued drifter motion (middle panel) much better resembles
the GASB (bottom panel) than the assessment based on drogued drifter
motion (top panel).  Similar behavior has been recently reported
by \citep{vanSebille-etal-21} from direct inspection of individual
trajectories produced by a reduced set of custom-made undrogued
drifters along with drogued drifters of the GDP type.  As opposed
to drogued GDP drifter motion, which mostly reflects water motion
at 15-m depth in the water column, undrogued drifter motion is
affected by ocean current shear between 0 and 15 m, and wind and
wave drag effects, mediated by ``inertial'' effects, i.e., those
due to the buoyancy, size, and shape of the drifters
\citep{Furnans-etal-08, Beron-etal-16, Beron-etal-19-PoF,
Olascoaga-etal-20, Miron-etal-20-GRL, Miron-etal-20-PoF, Beron-21-ND}.
The resemblance---impressively quite close---between the observed
GASB distribution (in the bottom panel of Fig.\ \ref{fig:gasb}) and
that suggested by undrogued drifter trajectories indicates a dominant
role of inertial effects in the drift of \sar.  Indeed, these appear
as an important mechanism for pelagic \sar{ }to inundate the IAS.

The rest of this note is devoted to go beyond the discussion of
direct probability density evolution with the twofold goal in mind
to gain insight into: \sar{ }connectivity as inferred by a Markov
chain constructed using undrogued drifter trajectories exclusively,
which show best agreement with actual \sar{ }movement  (Sec.\
\ref{sec:geo}); and how \sar{ }pathways connecting a potential
remote source with the IAS are accomplished in the most effective
manner (Sec.\ \ref{sec:tpt}).  This is done by applying specialized
probabilistic tools, which are briefly reviewed before the results
are discussed in each case. Section \ref{sec:dis} comments on aspects
of the movement of \sar{ }rafts that are needed to be accounted for
in order to make the most out of the learnings from the analysis
presented here. This brief communication is closed with a summary
(Sec.\ \ref{sec:sum}).

\section{Dynamical geography of \sar}\label{sec:geo}

Time-asymptotic aspects of the evolution along the Markov chain
resulting from the discretization are encoded in the spectral
properties of $\mathsf P$.  If every state of the chain (box of the
covering) is visited irrespective of the starting state (property
called \emph{irreducibility} or \emph{ergodicity}, and formally
represented as $\forall\,i,j:b_i,b_j\in\mathscr D,\,\exists n_{ij}
\in\mathbb Z_0^+\setminus\{\infty\}:  \smash{(P^{n_{ij}})_{ij} >
0}$) and no state is revisited cyclically (property referred to as
\emph{aperiodicty} or \emph{mixing} and expressed as $\exists
i:b_i\in\mathscr D : \gcd\{n\in\mathbb Z_0^+ : \smash{(P^n)_{ii} >
0}\} = 1$), then the eigenvalue $\lambda = 1$ of $\mathsf P$ is the
largest possible and has multiplicity one.

The left eigenvector, $\boldsymbol\pi$, corresponding to $\lambda
= 1$ is strictly (componentwise) positive and represents an invariant
distribution (i.e., $\boldsymbol\pi = \boldsymbol\pi \mathsf P$),
which typically reveals local maxima (bumps) where trajectories
settle on in the long run.  Normalizing $\boldsymbol\pi$ to a
probability vector (i.e., such that $\sum_{i:b_i\in\mathscr D}\pi_i
= 1$) it follows that $\boldsymbol\pi = \mathbf p \mathsf P^\infty$
for any probability vector $\mathbf p$.   In other words,
$\boldsymbol\pi$ is a limiting or \defn{stationary distribution}
that time-asymptotically sets $\Pr(X_n \in b_i) = \pi_i$.

The right eigenvector corresponding to $\lambda = 1$ is a (column)
vector of ones, denoted $\mathbf 1$. Indeed $\mathsf P\mathbf 1 =
\mathbf 1$ due to stochasticity of $\mathsf P$, viz., $\sum_{j:b_j\in
\mathscr D} P_{ij} = 1$ for all $i$, trivially indicating a
\defn{backward-time basin of attraction} for the time-asymptotic
attractors where $\boldsymbol\pi$ maximizes: any probability vector
with support in $\mathscr D$ distributes in the long run like
$\boldsymbol\pi$.

\begin{remark}
  The irreducibility/ergodicity and aperiodicity/mixing conditions
  can only be met when the flow domain $\mathscr D$ is closed, which
  is not our case.  Our transition matrix is substochastic, namely,
  $\sum_{j:b_j\in\mathscr D} P_{ij} < 1$ for some $i$, which must
  be appropriately ``stochasticized,'' for which there are several
  options \citep{Froyland-etal-14, Froyland-etal-14c, Miron-etal-17}.
  Here we follow \citep{Miron-etal-21-Chaos} and augment the chain
  by appending what has been coined by \citep{Miron-etal-21-Chaos}
  a \defn{two-way-nirvana state}.  This is a virtual state that is
  appended to the chain in such a way that it compensates for any
  probability mass imbalances due to the openness of $\mathscr D$
  by reinjecting them back into the chain using available trajectory
  information.  We express this as follows. The transition matrix
  $\mathsf P$ computed from trajectories flowing in and out of
  $\mathscr D$, call it $\mathsf P^{\mathscr D}$, is replaced by a
  stochastic transition matrix $\mathsf P\in \mathbb R^ {(N+1)\times
  (N+1)}$ defined by
  \begin{equation}
	\mathsf P :=
	\begin{pmatrix}
		\mathsf P^{\mathscr D} & \mathbf p^{\mathscr D\to\omega}\\ 
		\mathbf p^{\mathscr D\leftarrow\omega} & 0
	\end{pmatrix}.
	\label{eq:closure}
  \end{equation}
  Here, $\omega$ is the two-way nirvana state alluded to above,
  used to augment the chain defined by $\mathsf P$. In turn, vector
  $\smash{\mathbf p^{\mathscr D\to\omega} := (1 - \sum_{j :
  b_j\in\mathscr D} P_{ij}) \in \mathbb R^{N \times 1}}$ gives the
  outflow from $\mathscr D$ and the probability vector $\smash{\mathbf
  p^{\mathscr D\leftarrow\omega}\in \mathbb R^{1\times N}}$ gives
  the inflow, which is estimated from the trajectory data.
  \label{rem:closure}
\end{remark}

Regions where left eigenvectors of $\mathsf P$ with $\lambda < 1$
locally maximize indicate \defn{almost-invariant attracting sets}.
Plateaus in the corresponding right eigenvectors highlight basins
of attraction or regions where trajectories temporarily converging
in the almost-invariant attractors begin.  This imposes constraints
on connectivity \citep{Dellnitz-Hohmann-97, Dellnitz-Junge-99,
Froyland-Dellnitz-03, Koltai-10}.  The collection of nonoverlapping
basins of attraction form a partition of the dynamics into the
weakly (dis)connected ``provinces'' of what has been termed
\citep{Froyland-etal-14, Miron-etal-17} a \defn{dynamical geography},
i.e., one that does not depend on arbitrary geographic demarcations
but rather on the intrinsic flow characteristics of each province.

The eigenvector method has been applied to surface drifter and
submerged float trajectories suggesting a characterization of
preferred pollution centers in the subtropical gyres
\citep{Froyland-etal-14, Miron-etal-19-Chaos, Miron-etal-21-Chaos}
and in marginal seas, both at the surface \citep{Miron-etal-17} and
at depth \citep{Miron-etal-19-JPO}, as almost-invariant attracting
sets with corresponding basins of attraction spanning areas as large
as those of the corresponding geographic basins, suggesting a strong
influence of the regions collecting pollutants on their transport.

\begin{figure}[t!]
  \centering%
  \includegraphics[width=.65\textwidth]{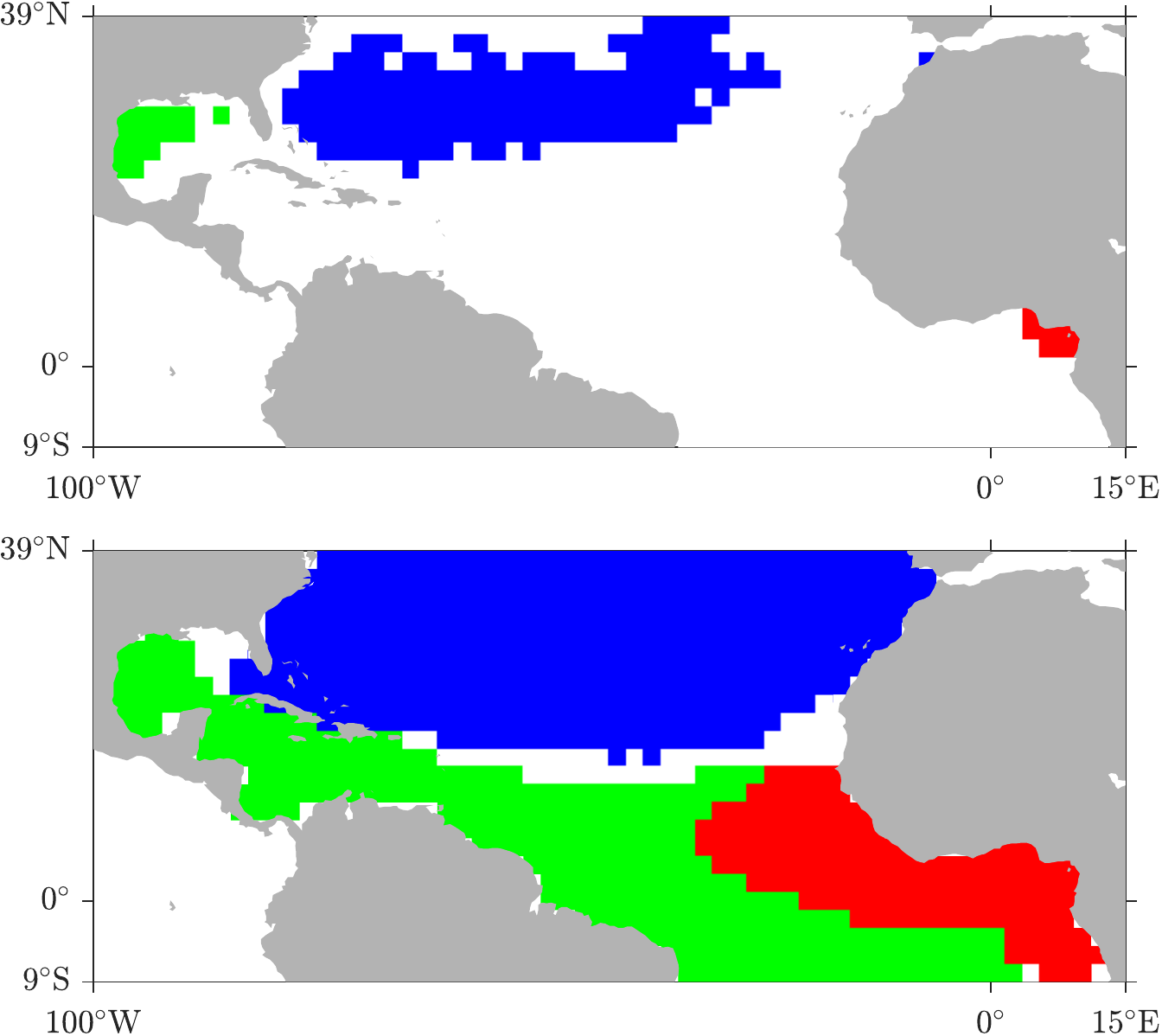} \caption{(top)
  Almost-invariant attracting sets inferred from the spectral
  analysis of a transition matrix representing conditional probabilities
  of undrogued GDP drifters from the NOAA database to move between
  boxes of a grid laid down on the Atlantic Ocean domain shown.
  (bottom) Basins of attraction for the attractors in the bottom
  panel, i.e., the regions where trajectories, which, ending in the
  long run in each of the like-colored patches on the top, begin.}
  \label{fig:geo}%
\end{figure}

Similar strong influence imposed by the northwestern Gulf of Mexico,
the Sargasso Sea, and the Gulf of Guinea is revealed from the
analysis of our $\mathsf P$, constructed using undrogued GDP drifter
trajectories from the NOAA dataset.  The top panel of Fig.\
\ref{fig:geo} shows the locations where first and second subdominant
left eigenvectors of $P$ maximize, which identify the aforementioned
geographical locations as almost-invariant attracting sets.  The
corresponding basins of attraction, as revealed by the first and
second subdominant right eigenvectors of $\mathsf P$, are shown in
the bottom panel of Fig.\ \ref{fig:geo}.  These are depicted with
the same color as the corresponding attractors to facilitate the
association. (In every case we have made use of the sparse eigenbasis
approximation \citep{Froyland-etal-19}, which enables multiple
feature extraction with application to coherent set identification.)
Note the large geographical regions covered by the basins, which
represent the domains of influence of the corresponding attractors.
For instance, trajectories passing through the Caribbean Sea on
their way into the northwestern Gulf of Mexico most likely start
anywhere in the Atlantic domain shown here except in the subtropical
gyre.  This suggests weak communication between the Sargasso Sea
and the GASB region, at least for the average undrogued GDP drifter
motion.

\section{Transition paths of \sar}\label{sec:tpt}

To unveil specific connectivity routes, additional probabilistic
tools are needed.  Particularly useful are those provided by the
\defn{Transition Path Theory} (\defn{TPT}) \citep{E-VandenEijnden-06,
Metzner-etal-09}, which has been recently adapted to open dynamical
systems \citep{Miron-etal-21-Chaos}.

\begin{figure}[t!]
  \centering%
  \includegraphics[width=.35\textwidth]{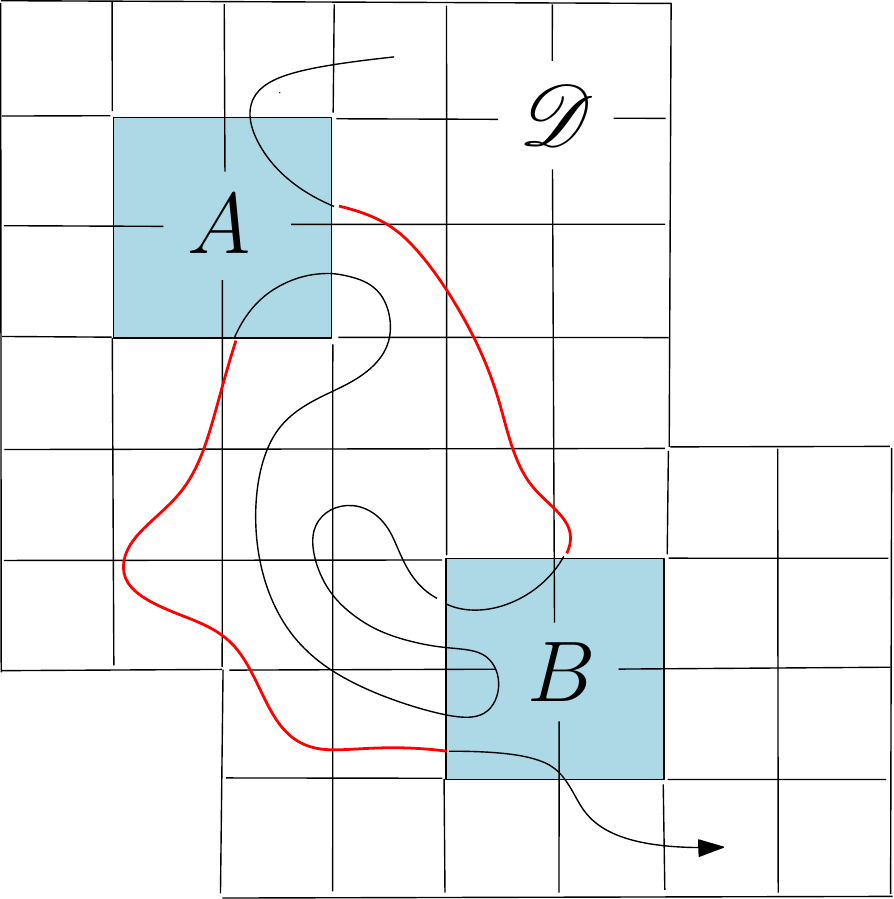}%
  \caption{Schematic representation of a piece of a hypothetical
  infinitely long drifter trajectory (black) that densely, albeit
  not necessarily uniformly, fills a closed flow domain $\mathscr
  D$, partitioned into boxes (black). Indicated are disconnected
  source ($A\subset \mathscr D$) and target ($B\subset \mathscr D$)
  sets. Highlighted in red are two members of an ensemble of reactive
  trajectories.  These are the trajectory subpieces that connect
  $A$ with $B$ in direct transition from $A$ to $B$, i.e., without
  returning back to $A$ or going through $B$ in between. The
  Transition Path Theory (TPT) provides a statistical characterization
  of the ensemble of such reactive trajectory pieces, highlighting
  the dominant communication conduits or transition paths between
  $A$ to $B$.}
  \label{fig:tpt-cartoon}%
\end{figure}

Developed to investigate rare events in complex systems, such as
chemical reactions or conformation changes of molecules, TPT provides
a statistical characterization of ensembles of ``reactive''
trajectories, namely, trajectories along which direct transitions
between a source set (or ``reactant'') $A$ and a target set (or
``product'') $B$ in a Markov chain take place (Fig.\
\ref{fig:tpt-cartoon}).

Applications of TPT have now gone beyond the study of molecular
dynamics or chemical kinetics\citep{Noe-etal-09, Metzner-etal-06,
Metzner-etal-09, Meng-etal-16, Thiede-etal-19, Liu-etal-19,
Strahan-etal-21}. TPT has been used to shed light on pollution
pathways in the ocean \citep{Miron-etal-21-Chaos}, paths of the
upper \citep{Drouin-etal-22} and lower \citep{Miron-etal-22} limbs
of the overturning circulation in the Atlantic Ocean, and even
atmospheric phenomena such as sudden stratospheric warmings
\citep{Finkel-etal-20}.

The main objects of TPT are the \emph{forward}, $\mathbf q^+ =
(q^+_i)\in \mathbb R^{1\times N}$, and \emph{backward}, $\mathbf
q^- = (q^-_i)\in \mathbb R^{1\times N}$, \defn{committor probabilities}.
These give the probability of a random walker initially in $b_i$
to first enter $B$ and last exit $A$, respectively. The committors
are fully computable from $\mathsf P$ and $\boldsymbol\pi$, according
to:
\begin{equation}
   \left\{
   \begin{aligned}
   &\restr{\mathbf q^\pm}{\mathcal D \setminus (A\cup B)} = \restr{\mathsf P^\pm}{\mathcal D \setminus A\cup B,\mathcal D} \mathbf q^\pm,\\
   &\restr{\mathbf q^+}{A}  = \mathbf 0^{|A|\times 1},\quad \restr{\mathbf q^-}{B} = \mathbf 0^{|B|\times 1},,\\
   &\restr{\mathbf q^+}{B}  = \mathbf 1^{|B|\times 1},\quad \restr{\mathbf q^-}{A} = \mathbf 1^{|A|\times 1}.
  \end{aligned}
  \right.
  \label{eq:q}
\end{equation}	
Here, $\mathsf P^+ = \mathsf P$; $P^-_{ij} := \Pr(X_n\in b_j\mid \ X_{n+1}\in b_i)
= \frac{\pi_j}{\pi_i}P_{ji}$ are the entries of the time-reversed
transition matrix, i.e., for the original chain $\{X_n\}$ traversed
in backward time, $\{X_{-n}\}$; and $\vert_S$ denotes restriction
to the subset $S$ while $\vert_{S,S'}$ that to the rows corresponding
to $S$ and columns to $S'$.

The committor probabilities are used to express several statistics
of the ensemble of reactive trajectories. The main ones are:
\begin{enumerate}
  \item The \defn{distribution of reactive trajectories}, $\boldsymbol
  \pi^{AB} = (\pi^{AB}_i)\in \mathbb R^{1\times N}$, where $\pi^{AB}_i$
  is defined as the joint probability that a trajectory is in box
  $b_i$ while transitioning from $A$ to $B$, and is computable as
  \citep{Metzner-etal-09, Helfmann-etal-20}
  \begin{equation}
	 \pi^{AB}_i = q^-_i\pi_iq^+_i.
    \label{eq:mu}
   \end{equation}
   It describes the bottlenecks during the transitions, i.e., where
   reactive trajectories spend most of their time. Clearly,
   $\pi^{AB}_{i: b_i\in A\cup B} \equiv 0$.
	\item The \defn{effective current of reactive trajectories},
   $\mathsf f^+ = (f^+_{ij})\in \mathbb R^{N\times N}$, where
   $f^+_{ij}$ gives the net flux of trajectories going through
   $b_i$ at time $nT$ and $b_j$ at time $(n+1)T$ on their way from
   $A$ to $B$, indicates the most likely transition channels. This
   is computable according to \citep{Helfmann-etal-20}
   \begin{equation}
	   f^+_{ij} = \max\left\{f^{AB}_{ij} -
	   f^{AB}_{ji},0\right\},\quad
	   f^{AB}_{ij} = q^-_i\pi_iP_{ij}q^+_j.
	   \label{eq:f}
   \end{equation}
\end{enumerate}

\begin{remark}
  Actual paths between $A$ and $B$ are random, describing meanders
  and recirculations, but TPT concerns their average behavior,
  revealing the dominant transition channels from $A$ to $B$.  If
  $\smash{f^{AB}_{ij} \approx f^{AB}_{ji}}$, then a lot of reactive
  flux is going both ways, and one can expect actual trajectories
  to meander and loop until getting to $B$.  But if $\smash{f^{AB}_{ij}
  \gg f^{AB}_{ji}}$, then one can expect a more clear-cut flow from
  $A$ to $B$.  This is hard---if not impossible---to visualize from
  direct evolution of densities (under left multiplication by
  $\mathsf P$), and if the effective progression of probability
  mass from $A$ to $B$ is highly noisy, other TPT diagnostics such
  as the effective reactive time from $A$ to $B$ cannot be inferred
  from direct evolution.
\end{remark}
\begin{remark}
  An important additional observation regards TPT for open systems,
  as is the case of interest here with $\mathscr D$ representing
  an open flow domain. In such systems, transitions between $A$ and
  $B$ should be constrained to take place within $\mathscr D$, i.e.,
  avoiding going through $\omega$, the (virtual) two-way nirvana
  state appended to compensate for probability mass imbalances (cf.\
  Remark \ref{rem:closure}). This can be easily accomplished
  \citep{Miron-etal-21-Chaos} by replacing $\boldsymbol \pi$ in the
  TPT formulae above by $\boldsymbol\pi\vert_{\mathscr D}$, where
  $\boldsymbol\pi$ is the stationary distribution of the transition
  matrix $\mathsf P$ for the closed system on $\mathscr D\cup\omega$,
  and $\mathsf P$ (in the TPT formulae) by the substochastic
  transition matrix $\mathsf P^{\mathscr D}$.
\end{remark}

\begin{figure}[t!]
  \centering%
  \includegraphics[width=.65\textwidth]{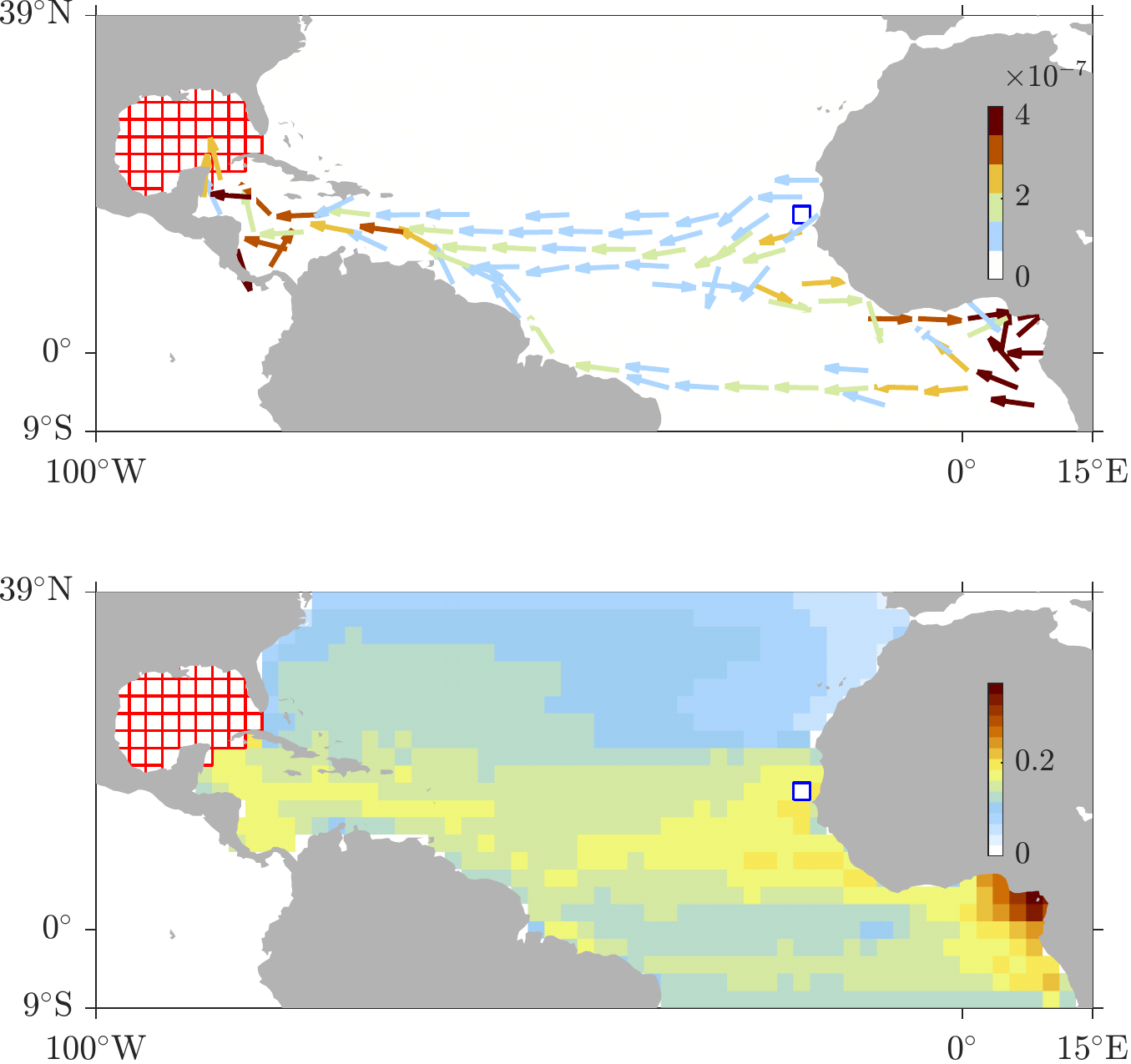}%
  \caption{(top panel) Currents of reactive trajectories between a
  hypothetical \sar{ }source off West Africa and target in the Gulf
  of Mexico, revealing the averaged, dominant pathways (transition
  paths) of \sar{ } into the IAS.  Colors represent vector magnitude.
  (bottom panel) Distribution of reactive trajectories revealing
  where the bottlenecks of the transitions take place.}
  \label{fig:tpt}%
\end{figure}

In the top panel of Fig.\ \ref{fig:tpt} we show the effective current
of reactive trajectories resulting between a hypothetical source
set $A$ off West Africa (blue box) and a target set $B$ formed by
the union of boxes covering the Gulf of Mexico (red boxes) when TPT
is applied on the Markov chain constructed using undrogued drifter
trajectories.  To visualize $\mathsf f^+$, we follow
\citep{Helfmann-etal-20, Miron-etal-21-Chaos} and for each $b_i$
we estimate the vector of the average direction and the amount of
the effective reactive current to each $b_j$, $j\neq i$. The location
of $A$ has been intentionally chosen to lie in a region near the
triple boundary of the dynamical provinces (cf.\ Fig.\ \ref{fig:geo},
bottom panel), which minimizes the influence of the almost-invariant
forward attractors within each basin of attraction: a probability
density initialized there in principle has equal chance to converge,
temporarily, into any of each of the three almost-attracting sets.
On the other hand, placing $A$ off West Africa as chosen finds
rationale in the fact that upwelling-favorable winds there may
provide the required nutrients to trigger blooming by vertical
pumping them into the mixed layer.  Note that the eastern extent
of the GASB coincides roughly with this location (cf.\ Fig.\
\ref{fig:gasb}).  The bottom panel of Fig.\ \ref{fig:tpt} uncovers
the spatial locations of the bottleneck of the transition paths,
namely, where the paths spend most of their time in their direct
transition from the source into the target.

Note the two routes revealed in the top panel of Fig.\ \ref{fig:tpt}:
a direct route roughly along the GASB and another more southern
route, reported for the first time here, going through the Gulf of
Guinea and then passing by the mouth of the Amazon River.  The
southern path is strongly influenced by the almost-invariant
attracting set in the Gulf of Guinea, which makes the trajectories
bottleneck there. The latter is highlighted by the large values
taken by the distribution of reactive trajectories in the bottom
panel of Fig.\ \ref{fig:tpt}.  The distribution of reactive
trajectories maximize along the two transitions routes, the mouth
of the Amazon River, and of course the Caribbean Sea, where the
reactive currents clutter.

Two additional TPT diagnostics are:
\begin{enumerate}
  \item[3.] The \emph{rate of reactive trajectories} entering $B$
  (or, equivalently, exiting $A$), defined as the probability per
  time step of a reactive trajectory to enter $B$ (or exit $A$),
  which is computed as \citep{Metzner-etal-09, Helfmann-etal-20}
  \begin{equation}
	 k^{AB} = \sum_{i, j: b_i,b_j\in B} f^{AB}_{ij}.
  \label{eq:k}
  \end{equation}
  Dividing by the transition time step $T$, $k^{AB}$ has the
  interpretation of the frequency at which a trajectory enters $B$
  (or exists $A$) \citep{Miron-etal-21-Chaos}. 
  \item[4.] The \defn{expected duration} of a transition from $A$ to
  $B$, which is obtained by dividing the probability of a trajectory
  piece being reactive by the transition rate (interpreted as a
  frequency) \citep{Helfmann-etal-20}:
  \begin{equation}
  	 t^{AB} = \frac{\sum_{j:b_j\in\mathscr D\setminus A\cup B}
    \mu^{AB}_j}{k^{AB}}.
    \label{eq:t}
  \end{equation}
\end{enumerate}
The computed expected duration of the transition paths into the IAS
is about 12 yrs, which seems rather long.  But this should not
come as big surprise, given the strong influence exerted by the
Gulf of Guinea almost-attracting set. In fact, if the transitions
are set to avoid the basin of attraction of Gulf of Guinea attractor,
the expected duration drops down to 1.8 yrs or so.  

\begin{remark}
The above requires a modification to the TPT formulae, which is
achieved as follows \citep{Miron-etal-21-Chaos}.  If $S $, $S \cap
(A\cup B) = \emptyset$, is the set that wants to be avoided by the
reactive currents, then the forward committor, $\mathbf q^+$ in
\eqref{eq:q}, must be computed with $A$ replaced by $A\cup S$, while
the backward committor, $\mathbf q^-$ in \eqref{eq:q}, with $B$
replaced by $B\cup S$. With these replacements, the forward committor
now gives the probability to next transition to $B$ rather than to
$A$ or $S$ when starting in box $b_i$. In turn, the backward committor
gives the probability to have last come from $A$, not $B$ or $S$.
This way, the product of forward and backward committors becomes
the probability when initially in $b_i$ to have last come from $A$
and next go to $B$ while not passing through $A$, $B$, or $S$ in
between.
\end{remark}

\begin{figure}[t!]
  \centering%
  \includegraphics[width=.65\textwidth]{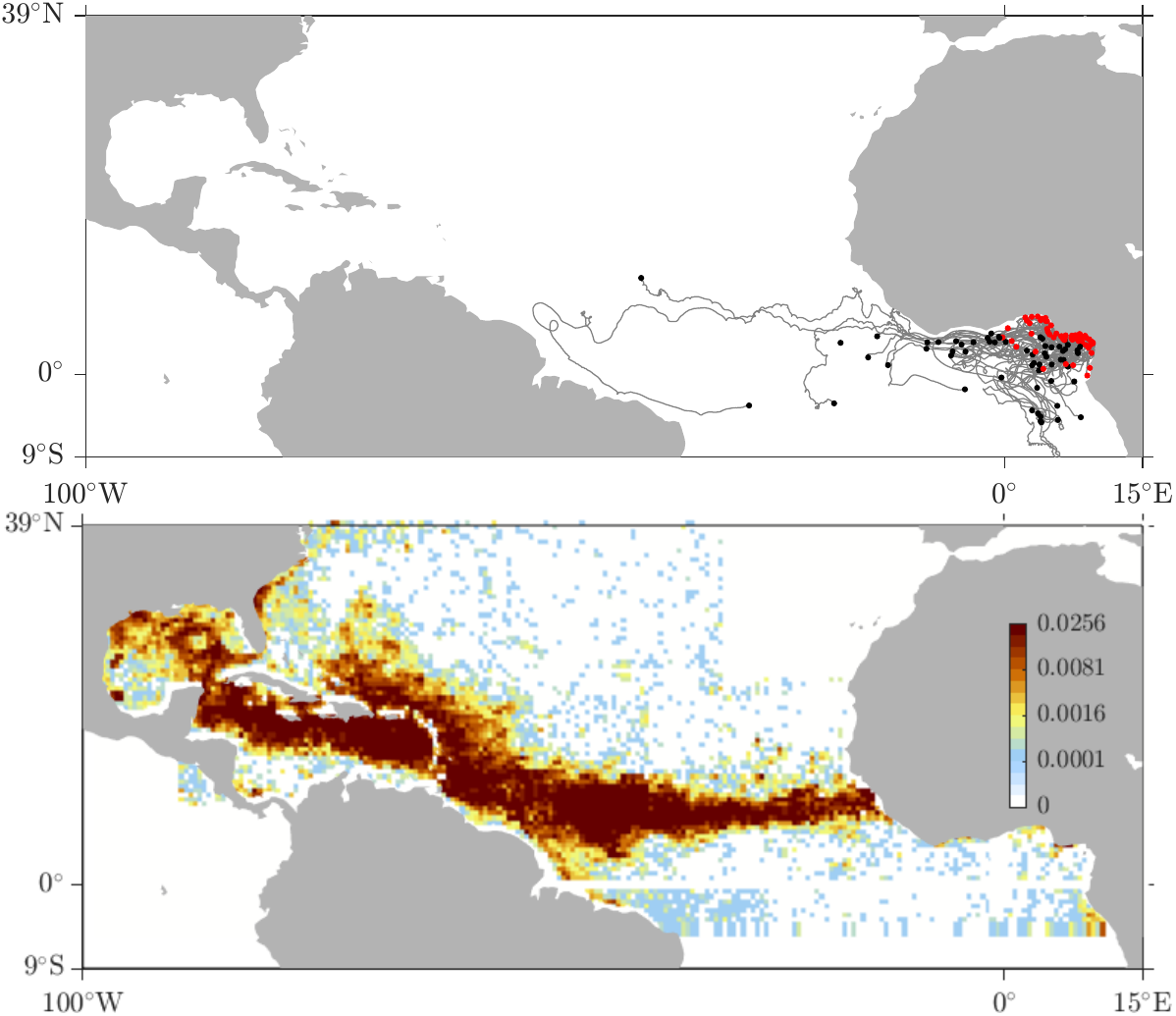}%
  \caption{(top) Trajectories of all undrogued GDP drifters from
  the NOAA dataset that converge in the Gulf of Guinea. Starting
  trajectory points are indicated by black dots, while endpoints
  by red dots. (bottom) Fourth-root transformation of maximum
  satellite-inferred percentage of \sar{ }coverage per pixel computed
  from boreal summer composites over 2011--2018.}
  \label{fig:guinea}%
\end{figure}

The reactive currents in the case that the basin of attraction of
the Gulf of Guinea attractor is avoided reveal a single route into
the IAS, roughly aligned along the GASB.  The physical significance
of this almost-invariant attractor is demonstrated in the top panel
Fig.\ \ref{fig:guinea}, which provides proof, independent of the
Markov chain model, of the existence of such an attractor.  The
figure shows individual trajectories of 64 undrogued GDP drifters
from the NOAA database over the period 1998--2014 which were found
to converge in the Gulf of Guinea.  A previous TPT analysis
\citep{Miron-etal-21-Chaos} had already identified the neighborhood
of the Gulf of Guinea as a province for a committor-based dynamical
geography for global marine debris circulation.  And it was also
pointed out \citep{Sutton-etal-17} that the Gulf of Guinea is a
mesopelagic niche with genomic characteristics that are different
than its surroundings.

The southern GASB branch unveiled by the TPT analysis is not an
artifact of it.  Support for the southern transition path is given
in the bottom panel Fig.\ \ref{fig:guinea}, which shows the fourth-root
transformation of the quantity shown in the bottom panel of Fig.\
\ref{fig:gasb} computed from boreal summer composites over 2011--2018.
Note the \sar{ }presence below the equatorial line, albeit in less
abundance than along the (main) GASB (branch).  This is consistent
with the TPT results (Fig.\ \ref{fig:tpt}), which reveals that the
southern path is weaker than the northern path, where the probability
of the reactive currents is stronger consistent with the higher
density of satellite-inferred floating \sar.

\section{Discussion}\label{sec:dis}

The analysis of the Markov chain model derived using undrogued
drifter trajectories indicates that the combined effects of winds,
waves, and surface currents can account for the basin-scale
distribution of the GASB. These results are in agreement with several
other studies that have suggested the importance of windage and
currents in shaping the trajectories and distribution of \sar{
}\citep{Brooks-etal-18, Putman-etal-20, Berline-etal-20}. Our
probabilistic analysis also indicates three, mostly independent,
oceanographic provinces associated with aggregations of \sar{ }that
occur in the Sargasso Sea, Gulf of Guinea, and Gulf of Mexico; a
finding consistent with \citep{Wang-etal-19}.  Furthermore, our
analysis suggests the existence of a secondary pathway by which
\sar{ }is transported from the tropical Atlantic into the Caribbean
Sea, south of the GASB (Figs.\ \ref{fig:tpt}--\ref{fig:guinea}).
While this pathway has not been previously described, evidence
consistent with this pathway is suggested in earlier studies.
Notably, \citep{Putman-etal-18} showed the importance of the Guinea
Current for moving \sar{ }along the coast of northern Brazil into
the Caribbean, which is fed by the South Equatorial Current. Satellite
imagery depicted in \citep{Putman-etal-18} and \citep{Gower-etal-13}
suggests that \sar{ }enters this region from south of the equator.
Thus, investigation of the ocean--atmosphere dynamics associated
with this newly described southern, slower pathway for \sar{ }and
the more northern, faster GASB \citep{Wang-etal-19}, could improve
the ability to predict basin-scale changes in \sar{ }distribution,
especially given very different time scales associated with the two
pathways.

For instance, it may be important to consider that the North
Equatorial Counter Current, running between 5 and 10$^{\circ}$N,
exhibits seasonal reversal \citep{Lumpkin-Garzoli-05}, which is not
taken into account by our autonomous TPT analysis, wherein the
transition matrix construction ignores the starting day of a drifter
trajectory.  To investigate the influence of the seasonal variability
in our results, a nonautonomous TPT analysis will be needed.  The
theoretical basis for such an analysis is now available
\citep{Helfmann-etal-20}, but still to be implemented.  A main
constraint in this case of interest here is imposed by the data
availability, which is unlikely to resist a seasonal partition.
Simulated trajectories may serve the purpose, but this is beyond
our scope here as is assessing interannual variability in \sar{
}connectivity,

Beyond the assumed time-homogeneity, the results of this work are
limited in two important counts.  First is the physiological changes
that \sar{ }rafts experience as they are dragged by the combined
action of ocean currents and winds, which the Markov chain model
constructed here did not account for.  A second aspect is the
architecture of the \sar{ }rafts, which plays a role in advection.
It is well-known that finite-size or ``inertial'' particles, even
neutrally buoyant ones, immersed in a fluid cannot adapt their
velocities to the carrying flow \citep{Maxey-Riley-83}.  Floating
particle motion is further complicated by the fact that the carrying
flow is not given by the ocean velocity below the sea surface itself,
but by the latter plus a buoyancy-dependent fraction of the wind
velocity right above the sea surface \citep{Beron-etal-19-PoF,
Olascoaga-etal-20, Miron-etal-20-GRL, Miron-etal-20-PoF, Beron-21-ND}.
This aspect is implicitly accounted for by our Markov chain model
as the undrogued drifters are affected by inertial effects. But
\sar{ }rafts are not isolated inertial particles.  Rather, they can
be more precisely envisioned as elastic networks of buoyant inertial
particles, with the buoyant inertial particles representing the
gas-filled bladders connected by flexible stems that keep the rafts
afloat. This was not represented by our Markov chain model. A
mathematical model has been recently proposed \citep{Beron-Miron-20}
and successfully tested \citep{Andrade-etal-22} against observations.
This model coupled with a physiological model of the transformations
that a \sar{ }raft experiences as it travels across regions with
varying environmental conditions would likely further elucidate the
complex dynamics of \sar{ }invasions in the IAS. Such work could
further improve transport predictions by distinguishing between the
roles of physical and physiological processes on observed distributions.

\section{Summary}\label{sec:sum}

In this note, we have constructed a time-homogeneous Markov chain
from trajectories of undrogued drifters from the NOAA Global Drifter
Program. With this tool we found that probability can distribute
in a manner that resembled very closely the recently observed
recurrent belt of high \sar{ }density in the tropical Atlantic that
has been termed the \emph{Great Atlantic \sar{ }Belt} (\emph{GASB}).
A spectral analysis of the associated transition matrix was further
found to unveil a forward attracting almost-invariant set in the
northwestern Gulf of Mexico with a corresponding basin of attraction
disconnected from the Sargasso Sea, but including the nutrient-rich
regions at the outflow of the Amazon and Orinoco Rivers and also,
marginally, the upwelling system off the northern coast of west
Africa. This represents a data-based inference of potential remote
sources of \sar{ }recurrently invading the Intra-Americas Seas
(IAS). By further applying Transition Path Theory (TPT) on the
data-derived Markov chain model, two potential pathways for \sar{
}into the IAS from the upwelling system off the coast of Africa
were revealed.  One TPT-inferred pathway was found to take place
along the GASB.  The second pathway, much slower than the GASB, was
found to be more southern, first going through the Gulf of Guinea,
then across the Atlantic toward the mouth of the Amazon River, and
finally along the eastern South American margin.  Why satellite
imagery does not reveal an intense southern GASB branch needs to
be investigated in view of its potentially important consequences
for bloom stimulation by nutrients from river runoff.  This should
be done by accounting for the physiological transformations that
\sar{ }rafts are acted upon as they are transported by the combined
action of currents and winds, which is the subject of ongoing
research.

\section*{Acknowledgements}

We are grateful to Philippe Miron, Luzie Helfmann, and Peter Koltai
for the benefit of many helpful discussions relating time-asymptotic
almost-invariant sets and transition path theory.

\section*{Funding}

This work was supported by the University of Miami's Cooperative
Institute for Marine \& Atmospheric Studies and the National
Science Foundation grant OCE2148499.

\section*{Competing interest}

The authors declare no competing interest.

\section*{Authors' contributions}

F.J.B.V.\ wrote the paper, carried out the computations reported
in it, and constructed the figures, except for the top panel of
Fig.\ 5, which was constructed by M.J.O, and the bottom panels of
Figs.\ 1 and 5, which were constructed by J.T. M.J.O.\ and N.F.P.\
discussed with F.J.B.V.\ \sar{ }phenomenology and contributed to
interpret the results from the analyses carried out in the paper.
G.J.G.\ and R.L.\ provided discretionary funding to support the
work.  All authors read and discussed the results of the work, and
contributed to edits.

\section*{Data availability}

The NOAA Global Drifter Program dataset is openly available in
\href{https://www.aoml.noaa.gov/phod/gdp/index.php}{https://\allowbreak
www.\allowbreak aoml.\allowbreak noaa.\allowbreak gov/\allowbreak
phod/gdp/\allowbreak index.php} \citep{Lumpkin-Pazos-07}. The
floating algae density fields, are produced by USF and distributed
by SaWS at
\href{https://optics.marine.usf.edu/projects/saws.html}{https://\allowbreak
optics.marine.usf.edu/\allowbreak projects/\allowbreak saws.html}
\citep{Wang-Hu-16}.

\bibliographystyle{mybst}
\bibliography{fot}

\begin{thebibliography}{62}
\expandafter\ifx\csname natexlab\endcsname\relax\def\natexlab#1{#1}\fi

\bibitem[Andrade-Canto {\rm et~al.}(2022)Andrade-Canto, Beron-Vera, Goni,
  Karrasch, Olascoaga and Trinanes]{Andrade-etal-22}
{\rm Andrade-Canto, F., Beron-Vera, F.~J., Goni, G.~J., Karrasch, D.,
  Olascoaga, M.~J. and Trinanes, J.} (2022).  {Carriers of \emph{Sargassum} and
  mechanism for coastal inundation in the Caribbean Sea}. {\em Phys.\ Fluids\/}
  34, 016602.

\bibitem[Berline {\rm et~al.}(2020)Berline, Ody, Jouanno, Chevalier, Andre,
  Thibaut and Menard]{Berline-etal-20}
{\rm Berline, L., Ody, A., Jouanno, J., Chevalier, C., Andre, J.-M., Thibaut,
  T. and Menard, F.} (2020).  {Hindcasting the 2017 dispersal of
  \emph{Sargassum} algae in the Tropical North Atlantic}. {\em Marine Pollution
  Bulletin\/} 158, 111431.

\bibitem[Beron-Vera(2021)]{Beron-21-ND}
{\rm Beron-Vera, F.~J.} (2021).  {Nonlinear dynamics of inertial particles in
  the ocean: From drifters and floats to marine debris and \emph{Sargassum}}.
  {\em Nonlinear Dyn.\/} 103, 1--26.

\bibitem[Beron-Vera and Miron(2020)]{Beron-Miron-20}
{\rm Beron-Vera, F.~J. and Miron, P.} (2020).  A minimal {M}axey--{R}iley model
  for the drift of \emph{{S}argassum} rafts. {\em J. Fluid Mech.\/} 904, A8.

\bibitem[Beron-Vera {\rm et~al.}(2016)Beron-Vera, Olascoaga and
  Lumpkin]{Beron-etal-16}
{\rm Beron-Vera, F.~J., Olascoaga, M.~J. and Lumpkin, R.} (2016).
  Inertia-induced accumulation of flotsam in the subtropical gyres. {\em
  Geophys. Res. Lett.\/} 43, 12228--12233.

\bibitem[Beron-Vera {\rm et~al.}(2019)Beron-Vera, Olascoaga and
  Miron]{Beron-etal-19-PoF}
{\rm Beron-Vera, F.~J., Olascoaga, M.~J. and Miron, P.} (2019).  {Building a
  Maxey--Riley framework for surface ocean inertial particle dynamics}. {\em
  Phys. Fluids\/} 31, 096602.

\bibitem[Bertola {\rm et~al.}(2020)Bertola, Boehm, Putman, Xue, Robinson,
  Harris, Baldwin, Overcast and Hickerson]{Bertola-etal-20}
{\rm Bertola, L.~D., Boehm, J.~T., Putman, N.~F., Xue, A.~T., Robinson, J.~D.,
  Harris, S., Baldwin, C.~C., Overcast, I. and Hickerson, M.~J.} (2020).
  Asymmetrical gene flow in five co-distributed syngnathids explained by ocean
  currents and rafting propensity. {\em Proceedings of the Royal Society B\/}
  287, 20200657.

\bibitem[Brooks {\rm et~al.}(2018)Brooks, Coles, Hood and
  Gower]{Brooks-etal-18}
{\rm Brooks, M.~T., Coles, V.~J., Hood, R.~R. and Gower, J. F.~R.} (2018).
  {Factors controlling the seasonal distribution of pelagic \emph{Sargassum}}.
  {\em Mar. Ecol. Prog. Ser.\/} 599, 1--18.

\bibitem[Dellnitz and Hohmann(1997)]{Dellnitz-Hohmann-97}
{\rm Dellnitz, M. and Hohmann, A.} (1997).  A subdivision algorithm for the
  computation of unstable manifolds and global attractors. {\em Numerische
  Mathematik\/} 75, 293--317.

\bibitem[Dellnitz and Junge(1999)]{Dellnitz-Junge-99}
{\rm Dellnitz, M. and Junge, O.} (1999).  On the approximation of complicated
  dynamical behavior. {\em SIAM J. Numer. Anal.\/} 36, 491--515.

\bibitem[Drouin {\rm et~al.}(2022)Drouin, Lozier, Beron-Vera, Miron and
  Olascoaga]{Drouin-etal-22}
{\rm Drouin, K.~L., Lozier, M.~S., Beron-Vera, F.~J., Miron, P. and Olascoaga,
  M.~J.} (2022).  Surface pathways connecting the {South and North Atlantic
  Oceans}. {\em Geophysical Research Letters\/} 49~(1), e2021GL096646.

\bibitem[{E} and {Vanden-Eijnden}(2006)]{E-VandenEijnden-06}
{\rm {E}, W. and {Vanden-Eijnden}, E.} (2006).  Towards a theory of transition
  paths. {\em J. Stat. Phys.\/} 123, 503--623.

\bibitem[Finkel {\rm et~al.}(2020)Finkel, Abbot and Weare]{Finkel-etal-20}
{\rm Finkel, J., Abbot, D.~S. and Weare, J.} (2020).  Path properties of
  atmospheric transitions: {I}llustration with a low-order sudden stratospheric
  warming model. {\em Journal of the Atmospheric Sciences\/} 77, 2327 -- 2347.

\bibitem[Franks {\rm et~al.}(2016)Franks, Johnson and Ko]{Franks-etal-16}
{\rm Franks, J., Johnson, D. and Ko, D.} (2016).  {Pelagic \emph{Sargassum} in
  the tropical North Atlantic}. {\em Gulf Caribbean Res\/} 27, C6--11.

\bibitem[Froyland and Dellnitz(2003)]{Froyland-Dellnitz-03}
{\rm Froyland, G. and Dellnitz, M.} (2003).  Detecting and locating
  near-optimal almost-invariant sets and cycles. {\em SIAM J. Sci. Comput.\/}
  24, 1839--1863.

\bibitem[Froyland {\rm et~al.}(2014{\natexlab{{\rm a}}})Froyland, Pollett and
  Stuart]{Froyland-etal-14c}
{\rm Froyland, G., Pollett, P.~K. and Stuart, R.~M.} (2014{\natexlab{{\rm
  a}}}).  A closing scheme for finding almost-invariant sets in open dynamical
  systems. {\em Journal of Computational Dynamics\/} 1~(1), 135.

\bibitem[Froyland {\rm et~al.}(2019)Froyland, Rock and
  Sakellariou]{Froyland-etal-19}
{\rm Froyland, G., Rock, C.~P. and Sakellariou, K.} (2019).  {Sparse eigenbasis
  approximation: Multiple feature extraction across spatiotemporal scales with
  application to coherent set identification}. {\em Commun Nonlinear Sci Numer
  Simulat\/} 77, 81--107.

\bibitem[Froyland {\rm et~al.}(2014{\natexlab{{\rm b}}})Froyland, Stuart and
  {van Sebille}]{Froyland-etal-14}
{\rm Froyland, G., Stuart, R.~M. and {van Sebille}, E.} (2014{\natexlab{{\rm
  b}}}).  How well-connected is the surface of the global ocean? {\em Chaos\/}
  24, 033126.

\bibitem[Furnans {\rm et~al.}(2008)Furnans, Imberger and
  Hodges]{Furnans-etal-08}
{\rm Furnans, J., Imberger, J. and Hodges, B.~R.} (2008).  Including drag and
  inertia in drifter modelling. {\em Environmental Modelling \& Software\/} 23,
  714--728.

\bibitem[Gower {\rm et~al.}(2013)Gower, Young and King]{Gower-etal-13}
{\rm Gower, J., Young, E. and King, S.} (2013).  {Satellite images suggest a
  new \emph{Sargassum} source region in 2011}. {\em Remote Sensing Letters\/}
  4, 764--773.

\bibitem[Helfmann {\rm et~al.}(2020)Helfmann, Borrell, Sch\"utte and
  Koltai]{Helfmann-etal-20}
{\rm Helfmann, L., Borrell, E.~R., Sch\"utte, C. and Koltai, P.} (2020).
  Extending transition path theory: {P}eriodically driven and finite-time
  dynamics. {\em J. Nonlinear Sci.\/} 30, 3321--3366.

\bibitem[Johns {\rm et~al.}(2020)Johns, Lumpkin, Putman, Smith, Muller-Karger,
  Rueda-Roa, Hu, Wang, Brooks, Gramer and Werner]{Johns-etal-20}
{\rm Johns, E.~M., Lumpkin, R., Putman, N.~F., Smith, R.~H., Muller-Karger,
  F.~E., Rueda-Roa, D.~T., Hu, C., Wang, M., Brooks, M.~T., Gramer, L.~J. and
  Werner, F.~E.} (2020).  The establishment of a pelagic sargassum population
  in the tropical atlantic: Biological consequences of a basin-scale long
  distance dispersal event. {\em Progress in Oceanography\/} 182, 102269.

\bibitem[Jouanno {\rm et~al.}(2021{\natexlab{{\rm a}}})Jouanno, Benshila,
  Berline, Souli\'e, Radenac, Morvan, Diaz, Sheinbaum, Chevalier, Thibaut,
  Changeux, Menard, Berthet, Aumont, Eth\'e, Nabat and
  Mallet]{Jouanno-etal-21b}
{\rm Jouanno, J., Benshila, R., Berline, L., Souli\'e, A., Radenac, M.-H.,
  Morvan, G., Diaz, F., Sheinbaum, J., Chevalier, C., Thibaut, T., Changeux,
  T., Menard, F., Berthet, S., Aumont, O., Eth\'e, C., Nabat, P. and Mallet,
  M.} (2021{\natexlab{{\rm a}}}).  {A NEMO-based model of \emph{Sargassum}
  distribution in the tropical Atlantic: description of the model and
  sensitivity analysis (NEMO-Sarg1.0)}. {\em Geoscientific Model Development\/}
  14~(6), 4069--4086.

\bibitem[Jouanno {\rm et~al.}(2021{\natexlab{{\rm b}}})Jouanno, Moquet,
  Berline, Radenac, Santini, Changeux, Thibaut, Podlejski, M{\'{e}}nard,
  Martinez, Aumont, Sheinbaum, Filizola and N'Kaya]{Jouanno-etal-21a}
{\rm Jouanno, J., Moquet, J.-S., Berline, L., Radenac, M.-H., Santini, W.,
  Changeux, T., Thibaut, T., Podlejski, W., M{\'{e}}nard, F., Martinez, J.-M.,
  Aumont, O., Sheinbaum, J., Filizola, N. and N'Kaya, G. D.~M.}
  (2021{\natexlab{{\rm b}}}).  Evolution of the riverine nutrient export to the
  tropical atlantic over the last 15 years: is there a link with sargassum
  proliferation? {\em Environmental Research Letters\/} 16, 034042.

\bibitem[Koltai(2010)]{Koltai-10}
{\rm Koltai, P.} (2010).  Efficient approximation methods for the global
  long-term behavior of dynamical systems -- theory, algorithms and examples.
  PhD thesis, Technical University of Munich.

\bibitem[LaCasce(2008)]{LaCasce-08}
{\rm LaCasce, J.~H.} (2008).  {Statistics from Lagrangian observations}. {\em
  Progr. Oceanogr.\/} 77, 1--29.

\bibitem[Lapointe {\rm et~al.}(2021)Lapointe, Brewton, Herren, Wang, amd
  D.~J.~{McGillicuddy Jr.}, Lindell, Hernandez and Morton]{LaPointe-etal-21}
{\rm Lapointe, B.~E., Brewton, R.~A., Herren, L.~W., Wang, M., amd
  D.~J.~{McGillicuddy Jr.}, C.~H., Lindell, S., Hernandez, F.~J. and Morton,
  P.~L.} (2021).  {Nutrient content and stoichiometry of pelagic Sargassum
  reflects increasing nitrogen availability in the Atlantic Basin}. {\em Nature
  Comms.\/} 12, 3060.

\bibitem[Lasota and Mackey(1994)]{Lasota-Mackey-94}
{\rm Lasota, A. and Mackey, M.~C.} (1994).  {\em Chaos, Fractals and Noise:
  Stochastic Aspects of Dynamics\/}, 2nd edn., vol.~97 of {\em Applied
  Mathematical Sciences\/}. New York: Springer.

\bibitem[Liu {\rm et~al.}(2019)Liu, Hickey, Minteer, Dickson and {Calabrese
  Barton}]{Liu-etal-19}
{\rm Liu, Y., Hickey, D.~P., Minteer, S.~D., Dickson, A. and {Calabrese
  Barton}, S.} (2019).  Markov-state transition path analysis of electrostatic
  channeling. {\em The Journal of Physical Chemistry C\/} 123, 15284--15292.

\bibitem[Lumpkin and Garzoli(2005)]{Lumpkin-Garzoli-05}
{\rm Lumpkin, R. and Garzoli, S.} (2005).  Near-surface circulation in the
  tropical atlantic ocean. {\em Deep-Sea Research I\/} 52, 495--518.

\bibitem[{Lumpkin} {\rm et~al.}(2012){Lumpkin}, {Grodsky}, {Centurioni}, {Rio},
  {Carton} and {Lee}]{Lumpkin-etal-12}
{\rm {Lumpkin}, R., {Grodsky}, S.~A., {Centurioni}, L., {Rio}, M.-H., {Carton},
  J.~A. and {Lee}, D.} (2012).  Removing spurious low-frequency variability in
  drifter velocities. {\em J. Atm. Oce. Tech.\/} 30, 353--360.

\bibitem[Lumpkin and Pazos(2007)]{Lumpkin-Pazos-07}
{\rm Lumpkin, R. and Pazos, M.} (2007).  {Measuring surface currents with
  Surface Velocity Program drifters: the instrument, its data and some recent
  results}. In {\em Lagrangian Analysis and Prediction of Coastal and Ocean
  Dynamics\/} (ed. A.~Griffa, A.~D. Kirwan, A.~Mariano, T.~\"{O}zg\"{o}kmen and
  T.~Rossby), chap.~2: pp. 39--67. Cambridge University Press.

\bibitem[Maxey and Riley(1983)]{Maxey-Riley-83}
{\rm Maxey, M.~R. and Riley, J.~J.} (1983).  Equation of motion for a small
  rigid sphere in a nonuniform flow. {\em Phys. Fluids\/} 26, 883.

\bibitem[Maximenko {\rm et~al.}(2012)Maximenko, Hafner and
  Niiler]{Maximenko-etal-12}
{\rm Maximenko, A.~N., Hafner, J. and Niiler, P.} (2012).  {Pathways of marine
  debris derived from trajectories of Lagrangian drifters}. {\em Mar. Pollut.
  Bull.\/} 65, 51--62.

\bibitem[McAdam and van Sebille(2018)]{McAdam-vanSebille-18}
{\rm McAdam, R. and van Sebille, E.} (2018).  Surface connectivity and
  interocean exchanges from drifter-based transition matrices. {\em Journal of
  Geophysical Research: Oceans\/} 123, 514--532.

\bibitem[Meng {\rm et~al.}(2016)Meng, Shukla, Pande and Roux]{Meng-etal-16}
{\rm Meng, Y., Shukla, D., Pande, V.~S. and Roux, B.} (2016).  Transition path
  theory analysis of c-src kinase activation. {\em Proceedings of the National
  Academy of Sciences\/} 113, 9193--9198.

\bibitem[Metzner {\rm et~al.}(2006)Metzner, {Sch\"utte} and
  {Vanden-Eijnden}]{Metzner-etal-06}
{\rm Metzner, P., {Sch\"utte}, C. and {Vanden-Eijnden}, E.} (2006).
  Illustration of transition path theory on a collection of simple examples.
  {\em J. Chem. Phys.\/} 125, 084110.

\bibitem[Metzner {\rm et~al.}(2009)Metzner, Sch\"utte and
  Vanden-Eijnden]{Metzner-etal-09}
{\rm Metzner, P., Sch\"utte, C. and Vanden-Eijnden, E.} (2009).  Transition
  path theory for {M}arkov jump processes. {\em Multiscale Modeling \&
  Simulation\/} 7, 1192--1219.

\bibitem[Miron {\rm et~al.}(2021)Miron, Beron-Vera, Helfmann and
  Koltai]{Miron-etal-21-Chaos}
{\rm Miron, P., Beron-Vera, F.~J., Helfmann, L. and Koltai, P.} (2021).
  Transition paths of marine debris and the stability of the garbage patches.
  {\em Chaos\/} 31, 033101.

\bibitem[Miron {\rm et~al.}(2022)Miron, Beron-Vera and
  Olascoaga]{Miron-etal-22}
{\rm Miron, P., Beron-Vera, F.~J. and Olascoaga, M.~J.} (2022).  {Transition
  paths of North Atlantic Deep Water}. {\em J. Atmos. Oce. Tech.\/} 39,
  959–971.

\bibitem[Miron {\rm et~al.}(2019{\natexlab{{\rm a}}})Miron, Beron-Vera,
  Olascoaga, Froyland, P\'erez-Brunius and Sheinbaum]{Miron-etal-19-JPO}
{\rm Miron, P., Beron-Vera, F.~J., Olascoaga, M.~J., Froyland, G.,
  P\'erez-Brunius, P. and Sheinbaum, J.} (2019{\natexlab{{\rm a}}}).
  {Lagrangian geography of the deep Gulf of Mexico}. {\em J. Phys. Oceanogr.\/}
  49, 269--290.

\bibitem[Miron {\rm et~al.}(2019{\natexlab{{\rm b}}})Miron, Beron-Vera,
  Olascoaga and Koltai]{Miron-etal-19-Chaos}
{\rm Miron, P., Beron-Vera, F.~J., Olascoaga, M.~J. and Koltai, P.}
  (2019{\natexlab{{\rm b}}}).  {Markov-chain-inspired search for MH370}. {\em
  Chaos: An Interdisciplinary Journal of Nonlinear Science\/} 29, 041105.

\bibitem[Miron {\rm et~al.}(2017)Miron, Beron-Vera, Olascoaga, Sheinbaum,
  P\'erez-Brunius and Froyland]{Miron-etal-17}
{\rm Miron, P., Beron-Vera, F.~J., Olascoaga, M.~J., Sheinbaum, J.,
  P\'erez-Brunius, P. and Froyland, G.} (2017).  {Lagrangian dynamical
  geography of the Gulf of Mexico}. {\em Scientific Reports\/} 7, 7021.

\bibitem[Miron {\rm et~al.}(2020{\natexlab{{\rm a}}})Miron, Medina, Olascaoaga
  and Beron-Vera]{Miron-etal-20-PoF}
{\rm Miron, P., Medina, S., Olascaoaga, M.~J. and Beron-Vera, F.~J.}
  (2020{\natexlab{{\rm a}}}).  {Laboratory verification of a Maxey--Riley
  theory for inertial ocean dynamics}. {\em Phys. Fluids\/} 32, 071703.

\bibitem[Miron {\rm et~al.}(2020{\natexlab{{\rm b}}})Miron, Olascoaga,
  Beron-Vera, {Tri\~nanes}, Putman, Lumpkin and Goni]{Miron-etal-20-GRL}
{\rm Miron, P., Olascoaga, M.~J., Beron-Vera, F.~J., {Tri\~nanes}, J., Putman,
  N.~F., Lumpkin, R. and Goni, G.~J.} (2020{\natexlab{{\rm b}}}).  {Clustering
  of marine-debris-and \emph{Sargassum}-like drifters explained by inertial
  particle dynamics}. {\em Geophys. Res. Lett.\/} 47, e2020GL089874.

\bibitem[No\'e {\rm et~al.}(2009)No\'e, Sch\"utte, Vanden-Eijnden, Reich and
  Weikl]{Noe-etal-09}
{\rm No\'e, F., Sch\"utte, C., Vanden-Eijnden, E., Reich, L. and Weikl, T.~R.}
  (2009).  Constructing the equilibrium ensemble of folding pathways from short
  off-equilibrium simulations. {\em Proceedings of the National Academy of
  Sciences\/} 106, 19011--19016.

\bibitem[Norris(1998)]{Norris-98}
{\rm Norris, J.} (1998).  {\em Markov Chains\/}. Cambridge University Press.

\bibitem[Olascoaga {\rm et~al.}(2020)Olascoaga, Beron-Vera, Miron,
  {Tri\~nanes}, Putman, Lumpkin and Goni]{Olascoaga-etal-20}
{\rm Olascoaga, M.~J., Beron-Vera, F.~J., Miron, P., {Tri\~nanes}, J., Putman,
  N.~F., Lumpkin, R. and Goni, G.~J.} (2020).  Observation and quantification
  of inertial effects on the drift of floating objects at the ocean surface.
  {\em Phys. Fluids\/} 32, 026601.

\bibitem[Olascoaga {\rm et~al.}(2018)Olascoaga, Miron, Paris, P\'erez-Brunius,
  P\'erez-Portela, Smith and Vaz]{Olascoaga-etal-18}
{\rm Olascoaga, M.~J., Miron, P., Paris, C., P\'erez-Brunius, P.,
  P\'erez-Portela, R., Smith, R.~H. and Vaz, A.} (2018).  {Connectivity of
  Pulley Ridge with remote locations as inferred from satellite-tracked drifter
  trajectories}. {\em Journal of Geophysical Research\/} 123, 5742--5750.

\bibitem[Paraguay-Delgado {\rm et~al.}(2020)Paraguay-Delgado, Carreno-Gallardo,
  Estrada-Guel, Zabala-Arceo, Martinez-Rodriguez and
  Lardizabal-Gutierre]{Paraguay-etal-20}
{\rm Paraguay-Delgado, F., Carreno-Gallardo, C., Estrada-Guel, I.,
  Zabala-Arceo, A., Martinez-Rodriguez, H.~A. and Lardizabal-Gutierre, D.}
  (2020).  {Pelagic \emph{Sargassum} spp. capture CO$_2$ and produce calcite}.
  {\em Environ Sci. Pollut. Res.\/} 42,
  https://doi.org/10.1007/s11356--020--08969--w.

\bibitem[Putman {\rm et~al.}(2018)Putman, Goni, Gramer, Hu, Johns, Trinanes and
  Wang]{Putman-etal-18}
{\rm Putman, N.~F., Goni, G.~J., Gramer, L.~J., Hu, C., Johns, E.~M., Trinanes,
  J. and Wang, M.} (2018).  {Simulating transport pathways of pelagic
  \emph{Sargassum} from the Equatorial Atlantic into the Caribbean Sea}. {\em
  Progress in Oceanography\/} 165, 205--214.

\bibitem[Putman {\rm et~al.}(2020)Putman, Lumpkin, Olascoaga, Trinanes and
  Goni]{Putman-etal-20}
{\rm Putman, N.~F., Lumpkin, R., Olascoaga, M.~J., Trinanes, J. and Goni,
  G.~J.} (2020).  {Improving transport predictions of pelagic
  \emph{Sargassum}}. {\em Journal of Experimental Marine Biology and Ecology\/}
  529, 151398.

\bibitem[Resiere {\rm et~al.}(2018)Resiere, Valentino, Neviere, Banydeen,
  Gueye, Florentin, Cabie, Lebrun, Megarbane, Guerrier and
  Mehdaoui]{Resiere-etal-18}
{\rm Resiere, D., Valentino, R., Neviere, R., Banydeen, R., Gueye, P.,
  Florentin, J., Cabie, A., Lebrun, T., Megarbane, B., Guerrier, G. and
  Mehdaoui, H.} (2018).  {\emph{Sargassum} seaweed on Caribbean islands: an
  international public health concern}. {\em The Lancet\/} 392, 2691.

\bibitem[van Sebille {\rm et~al.}(2021)van Sebille, Zettler, Wienders,
  Amaral-Zettler, Elipot and Lumpkin]{vanSebille-etal-21}
{\rm van Sebille, E., Zettler, E., Wienders, N., Amaral-Zettler, L., Elipot, S.
  and Lumpkin, R.} (2021).  Dispersion of surface drifters in the tropical
  atlantic. {\em Frontiers in Marine Science\/} 7, 607426.

\bibitem[Smetacek and Zingone(2013)]{Smetacek-Zingone-13}
{\rm Smetacek, V. and Zingone, A.} (2013).  Green and golden seaweed tides on
  the rise. {\em Nature\/} 504, 84--88.

\bibitem[Strahan {\rm et~al.}(2021)Strahan, Antoszewski, Lorpaiboon, Vani,
  Weare and Dinner]{Strahan-etal-21}
{\rm Strahan, J., Antoszewski, A., Lorpaiboon, C., Vani, B.~P., Weare, J. and
  Dinner, A.~R.} (2021).  Long- time-scale predictions from short-trajectory
  data: A benchmark analysis of the trp-cage miniprotein. {\em Journal of
  Chemical Theory and Computation\/} 17, 2948–2963.

\bibitem[Sutton {\rm et~al.}(2017)Sutton, Clark, Dunn, Halpin, Rogers,
  Guinotte, Bograd, Angel, Perez, Wishner, Haedrich, Lindsay, Drazen,
  Vereshchaka, Piatkowski, Morato, Blachowiak-Samolyk, Robison, Gjerde,
  Pierrot-Bults, Bernal, Reygondeau and Heino]{Sutton-etal-17}
{\rm Sutton, T.~T., Clark, M.~R., Dunn, D.~C., Halpin, P.~N., Rogers, A.~D.,
  Guinotte, J., Bograd, S.~J., Angel, M.~V., Perez, J. A.~A., Wishner, K.,
  Haedrich, R.~L., Lindsay, D.~J., Drazen, J.~C., Vereshchaka, A., Piatkowski,
  U., Morato, T., Blachowiak-Samolyk, K., Robison, B.~H., Gjerde, K.~M.,
  Pierrot-Bults, A., Bernal, P., Reygondeau, G. and Heino, M.} (2017).  A
  global biogeographic classification of the mesopelagic zone. {\em Deep Sea
  Research\/} 126, 85 -- 102.

\bibitem[Thiede {\rm et~al.}(2019)Thiede, Giannakis, Dinner and
  Weare]{Thiede-etal-19}
{\rm Thiede, E.~H., Giannakis, D., Dinner, A.~R. and Weare, J.} (2019).
  Galerkin approximation of dynamical quantities using trajectory data. {\em
  The Journal of Chemical Physics\/} 150, 244111.

\bibitem[Ulam(1960)]{Ulam-60}
{\rm Ulam, S.~M.} (1960).  {\em A Collection of Mathematical Problems\/}.
  Interscience.

\bibitem[{van Tussenbroek} {\rm et~al.}(2017){van Tussenbroek}, Arana,
  Rodriguez-Martinez, Espinoza-Avalos, Canizales-Flores, Gonzalez-Godoy,
  Barba-Santos, Vega-Zepeda and Collado-Vides]{vanTussenbroek-etal-17}
{\rm {van Tussenbroek}, B., Arana, H., Rodriguez-Martinez, R., Espinoza-Avalos,
  J., Canizales-Flores, H., Gonzalez-Godoy, C., Barba-Santos, M., Vega-Zepeda,
  A. and Collado-Vides, L.} (2017).  {Severe impacts of brown tides caused by
  \emph{Sargassum} spp. on near-shore Caribbean seagrass communities}. {\em
  Marine Pollution Bulletin\/} 122, 272--281.

\bibitem[Wang and Hu(2016)]{Wang-Hu-16}
{\rm Wang, M. and Hu, C.} (2016).  {Mapping and quantifying \emph{Sargassum}
  distribution and coverage in the Central West Atlantic using MODIS
  observations}. {\em Remote Sens. Environ.\/} 183, 350--367.

\bibitem[Wang {\rm et~al.}(2019)Wang, Hu, Barnes, Mitchum, Lapointe and
  Montoya]{Wang-etal-19}
{\rm Wang, M., Hu, C., Barnes, B., Mitchum, G., Lapointe, B. and Montoya,
  J.~P.} (2019).  {The Great Atlantic \emph{Sargassum} Belt}. {\em Science\/}
  365, 83--87.

\end{thebibliography}

\end{document}